\documentclass[letterpaper, 10 pt, conference]{ieeeconf}  
\IEEEoverridecommandlockouts                              

\usepackage[normalem]{ulem} 
\usepackage{cite}
\usepackage{amsmath,amssymb,amsfonts}
\usepackage{algorithmic}
\usepackage{graphicx}
\usepackage{textcomp}
\usepackage{xcolor}
\usepackage{caption}
\usepackage{amsmath}
\usepackage{amssymb}

\usepackage[nolist]{acronym} 
\usepackage{comment}
\usepackage{mathtools}
\usepackage{balance}
\usepackage{hyperref}

\usepackage{lipsum}
\usepackage{fancyhdr}

\fancypagestyle{sec}{\lhead{\tmpx}}

\fancypagestyle{arXiv}
{
   \fancyhf{}
   \chead{\textcolor{gray}{This paper has been accepted for publication at the \\
   IEEE International Conference on Robotics and Automation (ICRA), Paris, 2020}}
   \cfoot{\thepage}
   
}

\renewcommand{\phi}{\varphi}
\renewcommand{\epsilon}{\varepsilon}

\usepackage{hybridsystem}

\title{\LARGE \bf
Recurrent Neural Network Control of a Hybrid Dynamical Transfemoral Prosthesis with EdgeDRNN Accelerator
}

\author{Chang Gao$^{\star, 1}$, Rachel Gehlhar$^{\star, 2}$, Aaron D. Ames$^{2}$, Shih-Chii Liu$^{1}$ and Tobi Delbruck$^{1}$
\thanks{*This work was supported by the Samsung Global Research \textit{Neuromorphic Processor Project}, the National Science Foundation Graduate Research Fellowship under Grant No. DGE‐1745301 and NSF NRI Grant No. 1724464 and Swiss National Center of Competence in Research Robotics (NCCR Robotics).
This research was approved by California Institute of Technology Institutional Review Board with protocol no. 16-0693 for human subject testing. 
The authors also gratefully acknowledge discussions with Andrew Taylor and Prof.~Yisong Yue, 
and the opportunity to build the first prototype at the July 2019 \href{https://tellurideneuromorphic.org}{Telluride Neuromorphic Engineering Workshop}.}
\thanks{$^{\star}$The authors contributed equally to this work.}%
\thanks{$^{1}$Chang Gao, Shih-Chii Liu, Tobi Delbruck are with the Institute of Neuroinformatics, University of Zurich and ETH Zurich, Winterthurerstrasse 190, Switzerland
        {\tt\small chang@ini.uzh.ch, shih@ini.uzh.ch, tobi@ini.uzh.ch}}%
\thanks{$^{2}$Rachel Gehlhar and Aaron D. Ames are with the Department of Mechanical and Civil Engineering, California Institute of Technology, 1200 East California Boulevard, Pasadena, CA 91125, USA
        {\tt\small rgehlhar@caltech.edu, ames@caltech.edu}}%
}

\begin{document}
\maketitle
\thispagestyle{plain}
\pagestyle{plain}

\begin{abstract}

Lower leg prostheses could improve the life quality of amputees by increasing comfort and reducing energy to locomote, but currently control methods are limited in modulating behaviors based upon the human's experience. This paper describes the first steps toward learning complex controllers for dynamical robotic assistive devices. We provide the first example of behavioral cloning to control a powered transfemoral prostheses using a Gated Recurrent Unit (GRU) based recurrent neural network (RNN) running on a custom hardware accelerator that exploits temporal sparsity. The RNN is trained on data collected from the original prosthesis controller. The RNN inference is realized by a novel EdgeDRNN accelerator in real-time. Experimental results show that the RNN can replace the nominal PD controller to realize end-to-end control of the AMPRO3 prosthetic leg walking on flat ground and unforeseen slopes with comparable tracking accuracy. EdgeDRNN computes the RNN about 240 times faster than real time, opening the possibility of running larger networks for more complex tasks in the future. Implementing an RNN on this real-time dynamical system with impacts sets the ground work to incorporate other learned elements of the human-prosthesis system into prosthesis control.

\end{abstract}

\section{Introduction}
\thispagestyle{arXiv}
Even though there are over 222,000 transfemoral amputees in the United States \cite{LimbAmp}, the market for prostheses is mainly occupied by passive devices. 
These prostheses limit amputees' daily life by increasing their metabolic cost and constricting their locomotion abilities \cite{winter1991biomechanics}. 
Transfemoral amputees expend around 30\% more energy in walking compared to healthy humans \cite{EnergyExp}. 
Powered prostheses can reduce this metabolic cost and also increase their comfortable walking speed by providing net power to the user \cite{AntagActiveKnee, PowAnkleMetabolicHugh}.
This power is also particularly helpful for high energy tasks like stair-climbing \cite{PowAnkleFootStair, zhao2015realization}.
Motivated by this need, research on powered leg prostheses has largely focused on impedance control \cite{ImpCtrl, PowAnkleFootStair, DesignControlTransProsth, VirtConsCtrlProst}. 
Impedance control models each joint as a spring-damper system where the spring constant, damping coefficient, and equilibrium are parameters tuned by the user. Hence impedance control requires extensive tuning and is largely heuristic.
To address this limitation, \cite{aghasadeghi2013learning} developed a trajectory generation method for powered prostheses that is tracked by various online controllers, like PD control. The controller drives the motors' actual joint angles to the desired joint trajectories to yield human-like walking of the human-prosthesis system. However, the PD controller is very limited in its knowledge of the human-prosthesis system and hence its ability to control the prosthesis in response to the human.

\begin{figure}[t!]
	\centering
	\includegraphics[width=0.35\textwidth]{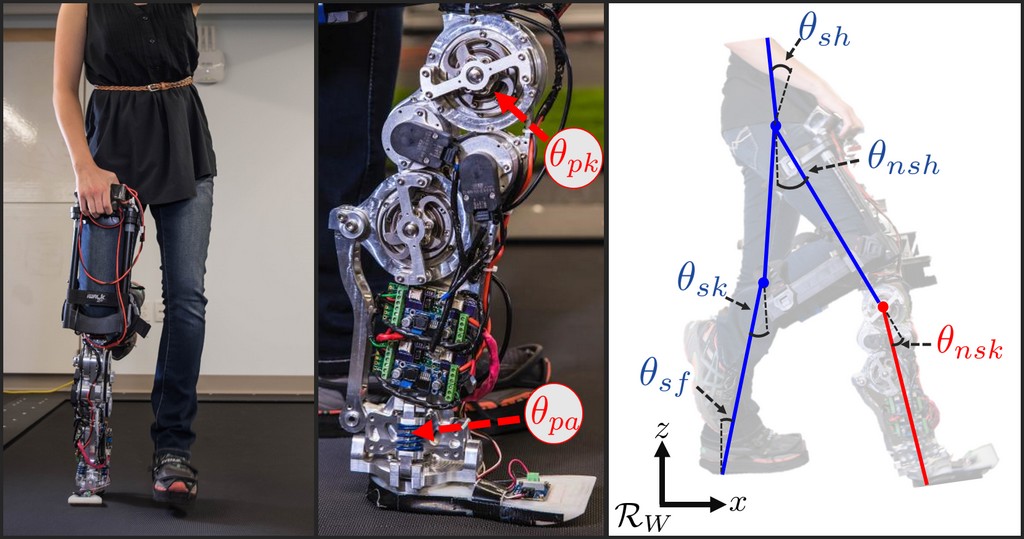}
	\caption{(Left) Human subject walking on powered transfemoral prosthesis, AMPRO3 (Middle). (Right) Robot model of human-prosthesis system labeled with joint angles.}
	\label{fig:ampro3}
	\vspace{-0.6cm}
\end{figure}

Motivated by the drawbacks of existing control methods, we investigate RNNs which have been used to understand human motion for use in prosthetic control systems. RNNs have been used with external sensors on a human to predict joint angles of human lower limbs \cite{PredictMoveRNN}, model the relation between arm muscles to finger flexion \cite{RNNMyoHand}, and estimate gait phase for a powered ankle exoskeleton. This ability to model complex human behavior draws us to discover how to integrate RNNs in real-time prosthesis control to open the door to using RNNs to decode patterns in human dynamics from sensor signals. In this paper we start this exploration by realizing RNNs for direct control of prostheses using imitation learning, a result absent in previous research, to serve as a proof-of-concept of RNN’s ability to be a part of real-time control of a hybrid dynamical system like the prosthesis.

Imitation learning has been used to reproduce human gestures on robots 
using hidden Markov model and Gaussian mixture regression~\cite{billard}. 
\cite{Evangelos} used a feed-forward deep neural network for online trajectory optimization with model predictive control in a continuous dynamical system. Imitation learning with high level motion primitives in non-agile hybrid systems
such as vision-based planning and control are popular, including~\cite{Finn2017,Li2017}. 
To the best of our knowledge, no research combines RNNs with imitation learning in a dynamical hybrid system with impacts. These system pose additional control challenges since the impacts can throw a robot off of its desired trajectory, requiring the controller to ensure rapid stabilization after impact. Also, the different domains of a hybrid system mean the controller has to work under varying conditions, e.g. with a ground reaction force present or not.
While control by an RNN can result in unprovable stability, the potential advantages
to having an RNN capable of performing real-time control on a prosthesis includes both building simpler models of complex controllers and dynamics for efficient computing, and capturing the nonlinearities and complexities of human walking behavior to include in prosthesis control methods.
In this study, we replaced the conventional PD controller with a gated-recurrent unit (GRU) RNN~\cite{gru_cho} using imitation learning. 
The RNN is trained on data collected during PD control of AMPRO3 walking on flat ground. 
An energy-efficient entry-level field programmable gate array (FPGA) accelerator called EdgeDRNN is used to run the RNN in real time. 
Experimental results show equivalent functionality between the RNN controller and the PD controller walking on flat ground and even on unseen slopes.
These results are demonstrated by experimental joint trajectory and torque data and by the supporting video~\cite{youtube}.
The main contributions of this paper are:
\begin{enumerate}
    \item First instance of imitation learning for hybrid dynamical systems with impacts;
    \item First realization of an end-to-end RNN control of a robotic prosthesis functioning with a human in the real-world using behavioral cloning;
    \item First work to use a hardware-accelerated RNN for real-time control.
\end{enumerate}
While the aim of this paper is not to improve upon the PD controller, by proving this RNN's ability to work on this dynamical system with impacts, we establish the first steps in using an RNN to improve upon existing prosthesis control methods. 
By learning other aspects of the human-prosthesis system, such as human comfort preferences \cite{PrefLearn}, various gait styles, and motion intent for transitions, we can bring the human into the loop of prosthesis control.
\section{Background}
This section introduces the framework of the traditional process of prosthesis controller generation that will be leveraged in the learning process, which is outlined subsequently.

\subsection{AMPRO with PD Controller}

We first present an overview of the methodology for synthesizing PD controllers for powered prostheses - specifically, AMPRO3 (Fig. \ref{fig:ampro3}), a powered transfemoral prosthesis developed at Caltech \cite{zhao2017preliminary}. In particular, this prosthesis with an actuated knee and ankle joint is modeled as a hybrid dynamical system. Desired outputs are defined that guide the behavior of the prosthesis wherein the parameters of these outputs are determined by an optimization problem that enforces impact invariance of these outputs \cite{ames2014human}. The end results are desired trajectories that are tracked via PD control on the device.

\newsec{Human-Prosthesis Model and Outputs.}
Trajectories are designed for the whole human-prosthesis system to yield stable walking. The system is modeled as a 5-link planar bipedal robot with configuration space $\mathcal{Q}_R: \theta = (\theta_{sf},\, \theta_{sk},\, \theta_{sh},\, \theta_{nsh}, \theta_{nsk})$, for the joint angles of the stance foot, stance knee, stance hip, non-stance hip, and non-stance knee, respectively, as shown in Fig. \ref{fig:ampro3} \cite{azimi2017robust}. 

To yield human-like walking, the control goal is to drive the actual robot outputs $y^a(\theta, \dot{\theta})$ to desired human outputs $y^d(t, \alpha)$.
These desired trajectories $y^d_2(t, \alpha)$ for the relative degree 2 (position) outputs are
defined by the canonical walking function \cite{ames2014human} with parameter set $\alpha$:
\begin{equation*}
    y^d_2(t, \alpha) = e^{-\alpha_4 t} ( \alpha_1 \cos(\alpha_2 t) + \alpha_3 \sin(\alpha_2 t) ) + \alpha_5. 
\end{equation*}
To develop a more robust controller, the trajectories are parameterized with a state-based phase variable:
\begin{equation*} 
    \rho(\theta) = \frac{\delta p_{hip}(\theta) - \delta p_{hip}^+}{v_{hip}}
\end{equation*}
in place of $t$ in the desired trajectories, $y^d_2(\rho(\theta), \alpha)$, 
where $\delta p_{hip}(\theta)$ is the linearized forward hip position, which increases linearly in a step cycle \cite{jiang2012outputs}. 
We define human-inspired outputs \cite{ames2014human}:
\begin{equation*}
    y(\theta, \dot{\theta}, \alpha) = 
    \begin{bmatrix}
    y_1(\theta, \dot{\theta}, \alpha) \\
    y_2(\theta, \alpha)
    \end{bmatrix}
    =
    \begin{bmatrix}
    y_1^a(\theta, \dot{\theta}) - v_{hip} \\
    y_2^a(\theta) - y_2^d(\rho(\theta), \alpha),
    \end{bmatrix}
\end{equation*}
where the relative degree one (velocity) output $y_1(\theta, \dot{\theta}, \rho, \alpha)$ is the difference between the actual and desired hip velocity (\textit{hip}), and the relative two outputs $y_2^a(\theta) - y_2^d(\rho, \alpha)$ are the differences between the actual and desired positions of the knee angles (\textit{sk, nsk}), non-stance slope (\textit{nsl}), and torso angle (\textit{tor}). For details, see \cite{zhao2014human}. 

\newsec{Gait Generation.}
To obtain human-like walking trajectories for $y_2^d(\rho(\theta), \alpha)$, we first model the system as a hybrid system \cite{ames2014human} to account for the discrete dynamics present at foot impact in walking. The impacts cause a discrete jump in the velocity of a walking robot model and could throw the system off of its desired trajectories. Hence, in our trajectory generation we enforce that the tracking of the relative degree~2 outputs remains invariant through impacts. These constraints, termed \textit{partial hybrid zero dynamics} constraints \cite{ames2014human}, are included in our optimization to solve for the parameters $\alpha$ and $v_{hip}$, which define our desired trajectories. Details can be found in \cite{zhao2017first}.

\begin{figure}[!t]
	\centering
	\includegraphics[width=0.27\textwidth]{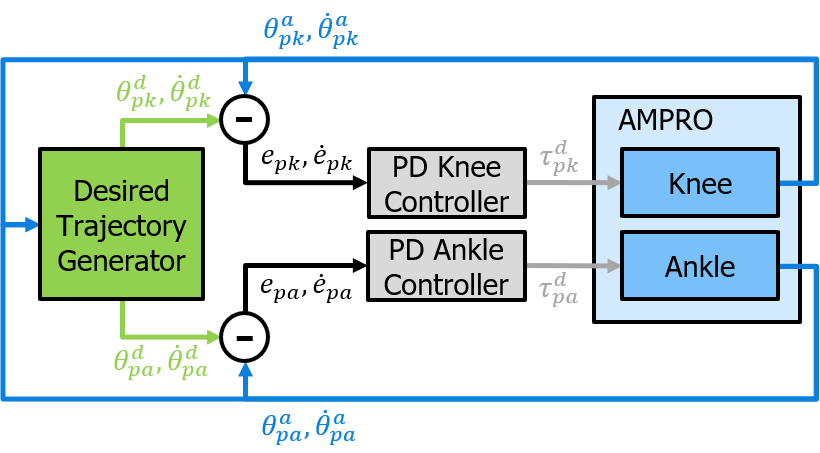}
	\caption{Architecture of the AMPRO PD control system.}
	\label{fig:pd_ctrl_sys}
    \vspace{-0.7cm}
\end{figure}

\newsec{Prosthesis Control.}
Fig.~\ref{fig:pd_ctrl_sys} shows the PD control architecture of the prosthesis. To apply a PD controller for trajectory tracking, the desired prosthesis knee angle $\theta^d_{pk}$ is determined from the desired trajectory $y^d_{2, sk}(\rho(\theta),  \alpha)$ for the prosthesis stance phase and $y^d_{2, nsk}(t, \alpha)$ for the prosthesis non-stance phase.
State-based control is used for the prosthesis stance phase to respond to the human's speed of progression in walking and direction of progression, i.e. forward or backward. This domain ends when $\rho(\theta) = \rho_{max}$, a constant determined by the user through testing.
Time-based control is used for prosthesis swing since the prosthesis does not have access to the human's stance hip position for the phase variable. Here $t$ is used in $y^d_{2, nsk}(t, \alpha$). This domain ends when $t = t_{max}$, a constant determined through optimization. For both phases $\theta^d_{pa}$ is set to 0 to act as a passive joint.

To have the actual prosthesis knee and ankle angles $(\theta^a_{pk},\, \theta^a_{pa})$ track these desired positions and velocities, a PD controller calculates the desired torque $\tau^d_{j}$ as follows:
\begin{equation*}
    \tau^d_{j} = K^{P}_{j}e_{j} + K^{D}_{j}\dot{e}_{j},
\end{equation*}
where $e_{j}=\theta^a_{j} - \theta^d_{j}$, $\dot{e}_{j}=\dot{\theta}^a_{j} - \dot{\theta}^d_{j}$ are respectively the errors between actual and desired angle and speed, and $j \in \{pk, pa\}$ indicates the joint. The PD gains $\{K^{P}_{j},\,  K^{D}_{j}\}|_{j  \in \{pk, pa\}}$ are selected such that the closed-loop linear output system is stable \cite{ames2014human}.

\subsection{Recurrent Neural Networks}
Recurrent neural networks (RNNs) are a family of artificial neural networks that are universal function approximators~\cite{ann}. 
RNNs have recurrent connections between their outputs and inputs and are a popular 
method in machine learning to model time series. 
State-of-the-art RNNs like the Gated-Recurrent Unit (GRU)~\cite{gru_cho} and Long Short-Term Memory (LSTM)~\cite{lstm_hoch97} models add gating units to the neurons. 
These units can overcome the vanishing gradient problem during training. 
These RNNs achieve high accuracy and are commercially deployed on many real-world tasks such as speech recognition and natural language processing. 
Outputs of a controller in a system can be seen as functions of sequential inputs; 
thus, RNNs are a candidate to approximate the temporal dependencies between outputs and inputs of the original controller.

\newsec{Gated-Recurrent Unit \& Delta Networks.}
In this work, a GRU-RNN based network is used to control AMPRO3. 
GRU-RNNs are computationally expensive and difficult to process on an embedded system in real time. 
Recurrent connections create data dependency between RNN inputs and outputs 
limiting the parallelism of this algorithm.
Thus, the memory bandwidth is the bottleneck of RNN computation 
on an embedded system. Modern von Neumann architecture based processors,
such as GPUs, suffer from low utilization rate of arithmetic units when running RNNs 
using a small batch size due to the insufficient 
memory bandwidth to external memory. 
Because of the poor power efficiency of GPUs on 
RNN inference with batch size of 1 of around 1.1\,GOp/s/W~\cite{Gao_iscas},
they are not suitable for a 
portable device with limited power budget like AMPRO3.

We proposed the bio-inspired DeltaGRU~\cite{neil2016delta} as an RNN 
variant that reduces the operations required for computing GRU-RNNs. 
Instead of multiplying weights with input and hidden activations as in GRU-RNNs, 
in DeltaGRU weights are multiplied with the the change of activations
between two adjacent time steps and then added to a memory term that 
is the accumulation of all previous products. 
By skipping the elements of a delta vector when their individual values 
are less than a defined threshold, 
the number of the corresponding matrix-vector multiply-and-accumulate (\textbf{MAC})
operations can typically be reduced by about a factor of 10 without loss of accuracy~\cite{neil2016delta}.


\newsec{EdgeDRNN Accelerator.}
The second generation DeltaRNN (DRNN) accelerator (based on~\cite{GaoDeltaRNN2018}) used in this paper is called EdgeDRNN~\cite{gao2019edgedrnn}. 
Like the original one, it skips unnecessary MAC operations by exploiting temporal sparsity like spiking neural networks, 
but it utilizes external DRAM memory for the large weight matrices.
This way it reduces the latency of the RNN inference and enables cheaper real-time 
inference. 
EdgeDRNN has 8 MAC units that support 16-bit fixed-point activations and 8-bit fixed-point weights. 
With a 125 MHz clock frequency, EdgeDRNN has a measured effective
throughput up to 20~GOp/s when running a large multi-layer DeltaGRU network with batch size 1, where weight reuse is not feasible~\cite{gao2019edgedrnn}.

\section{Controller Design}
This section outlines how the RNN controller is constructed with training data from the AMPRO3 prosthesis.

\subsection{RNN Controller}

\begin{figure}[!t]
	\centering
	\includegraphics[width=0.36\textwidth]{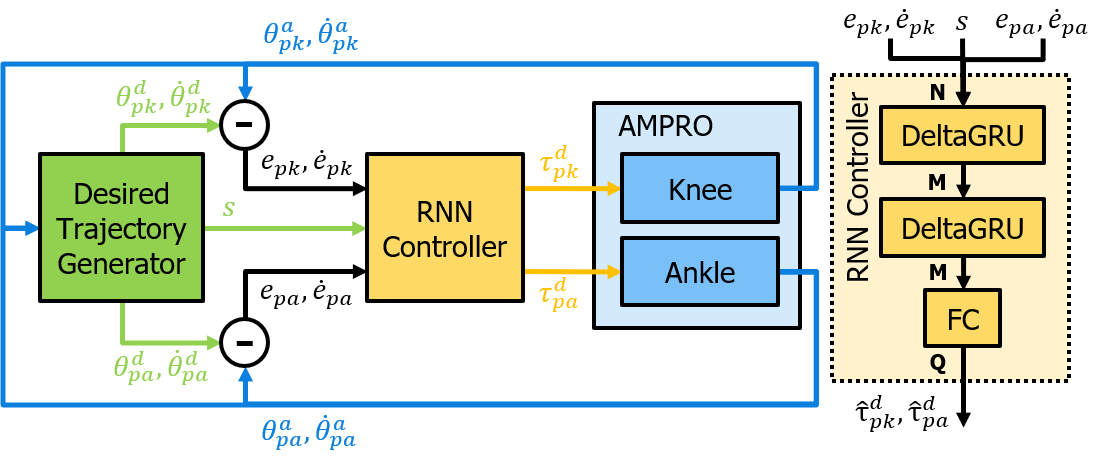}
	\caption{Architecture of the proposed AMPRO3 RNN controller system and the network structure.}
	\label{fig:rnn_ctrl_sys}
\end{figure}

Fig.~\ref{fig:rnn_ctrl_sys} shows the AMPRO3 control system architecture where the two original PD controllers for the knee and ankle are now replaced by the RNN controller.
An additional scalar input, $s$ to the RNN controller is the swing status calculated by the Desired Trajectory Model indicating whether the human leg is in swing phase or not. 
The model on the right of the figure shows 
the 2 GRU layers and 1 fully-connected (FC) layer of the RNN controller. 
Each GRU layer has $M$ neurons. The input dimension of the first GRU layer is $N=5$. 
The FC layer has $Q$ neurons to map the $M$-dimensional GRU-RNN output vector to a $Q$-dimensional vector, where $Q=2$ is the number of 
regression target variables and also the output vector dimension of the whole network. 
The following steps are required to deploy the RNN on AMPRO3:
\begin{enumerate}
    \item Collect data and labels from a demonstrator, which is the original PD controller. A user walks on AMPRO3 for an amount of time while the inputs and outputs of the demonstrator are recorded simultaneously. The recorded data 
    are divided into training, validation, and test sets. 
    \item Train the RNN controller network on the collected training set data and select network parameters using the validation set.
    \item Evaluate the performance of the RNN controller on the test set.
    \item Implement the RNN controller on the EdgeDRNN accelerator to control AMPRO3.
\end{enumerate}
\begin{figure}[!t]
	\centering
	\includegraphics[width=0.45\textwidth]{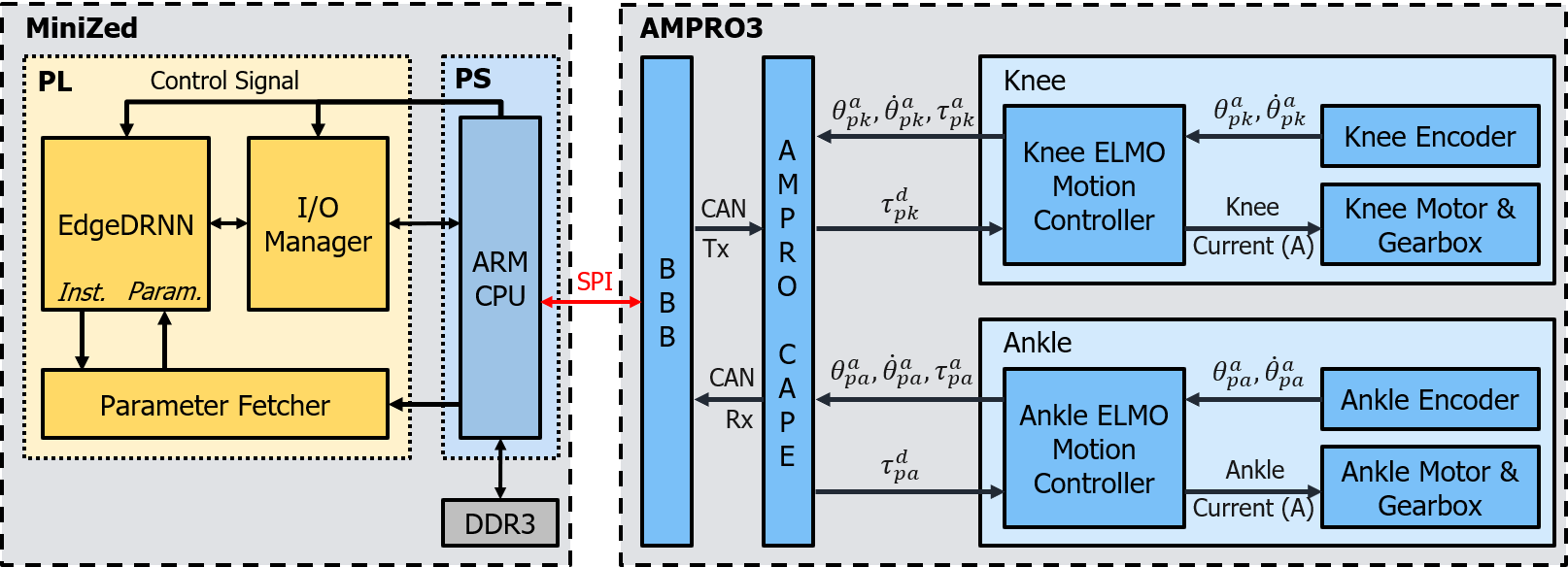}
	\caption{Block diagram of the hardware system with MiniZed and AMPRO3.}
	\label{fig:ampro_hardware}
	    \vspace{-0.6cm}
\end{figure}

\subsection{Data Collection \& Processing}
Data was collected by a subject walking with AMPRO3 on flat ground 5\,times. 
Each walk takes around 70\,seconds. 
All 5 walks give around 70,079 time steps of data in total with 200\,Hz sample rate. 
Each sample contains the desired and actual positions of the knee ($\theta^{d}_{pk}$, $\theta^{a}_{pk}$) and ankle ($\theta^{d}_{pa}$, $\theta^{a}_{pa}$), desired and actual velocities of the knee ($\dot{\theta}^{d}_{pk}$, $\dot{\theta}^{a}_{pk}$) and ankle ($\dot{\theta}^{d}_{pa}$, $\dot{\theta}^{a}_{pa}$), phase flag $s$ ($s=0$ for prosthesis non-stance, $s=1$ for prosthesis stance), and desired torques of the knee ($\dot{\tau}^{d}_{pk}$) and ankle ($\dot{\tau}^{d}_{pa}$). 
Errors between the desired and actual positions $e_{pk}, e_{pa}$ and velocities $\dot{e}_{pk}, \dot{e}_{pa}$ are calculated using aforementioned data. 
Three out of the five data files are used as the training set while both the validation and the test set have 1 data file each. 
Since the data are time series, we set a sequence length of $T$ time steps and stride of $1$ time step to select sequences out of each set. 
Training labels are given by $\tau_{dk}$ and $\tau_{da}$.

\subsection{Training Procedure}
The RNN controller network acts as a policy $\pi$ with parameters $\lambda$ that map inputs from the current and previous time steps to outputs of the current time, which is 
given as:
\begin{align*} 
    \mathbf{x}_{t} &= \left[ e_{pk,t}, e_{pa,t}, \dot{e}_{pk,t}, \dot{e}_{pa,t}, s_{t} \right], 
    \\  \mathbf{\hat{y}}_{t} &= \left[ \hat{\tau}^{d}_{pk,t}, \hat{\tau}^{d}_{pa,t} \right],
    \\  \mathbf{\hat{y}}_{t} &= \pi_{\lambda}\left( \mathbf{x}_{t}, \mathbf{\hat{y}}_{t-1}\right),
\end{align*}
where $\mathbf{x}_{t}$ and $\mathbf{\hat{y}}_{t}$ are the input and output vectors respectively at time step $t$. 
To train the network to infer the desired torques $\hat{\tau}^{d}_{pk,t}, \hat{\tau}^{d}_{pa,t}$ (i.e. a regression task), batches of training data are fed to the network for 
forward propagation. For each batch, the L1 loss is calculated as:
\begin{align*} 
    \ell^{i}_{t}\left(\mathbf{y}^i_t, \mathbf{\hat{y}}^i_t\right) &= \left| \mathbf{y}^i_t - \mathbf{\hat{y}}^i_t \right|,
    \\ 
    \mathcal{L} &= \frac{1}{B\times T}\sum^{B}_{i=1}\sum^{T}_{t=1}\ell^{i}_{t},
\end{align*}
where $i$ is the index of a sequence in a batch, $B$ the batch size, $\mathbf{y}^i_t=\left[\tau^{d,i}_{pk,t}, \tau^{d,i}_{pa,t}\right]$ the vector of labels, $\ell^i_{t}$ the loss of time step $t$, $\mathcal{L}$ the loss of a batch with sequences of length $T$.
The L1 loss is used instead of the L2 loss because of its robustness to outliers~\cite{imitation}. 


\section{Hardware Implementation}
\begin{figure}[!t]
	\centering
	\includegraphics[width=0.33\textwidth]{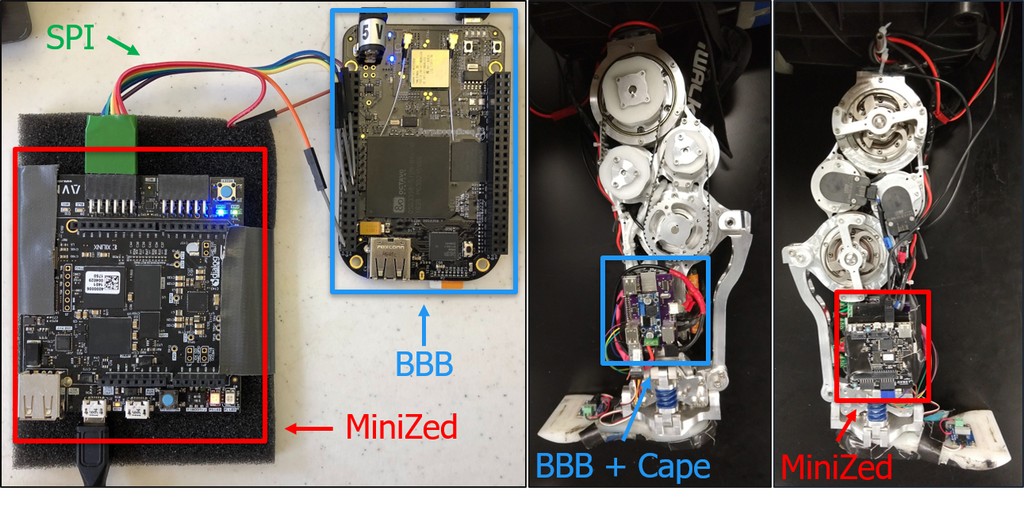}
	\caption{The MiniZed is attached to AMPRO3 and connected to the BBB through an SPI interface.}
	\label{fig:system_photo}
    \vspace{-0.7cm}
\end{figure}
The hardware of the original AMPRO3 and the EdgeDRNN systems and their integration will be covered in this section.

\subsection{AMPRO3}

Fig.~\ref{fig:ampro_hardware} shows the control architecture, including sensing and computation, of AMPRO3 \cite{zhao2017preliminary}.
It is run on a BeagleBone Black (\textbf{BBB}) micro-controller on-board the prosthetic. 
The BBB controller is coded in C++ and runs in a Robot Operating System at 200\,Hz.
A custom printed circuit board called AMPRO Cape for the BBB includes a Controller Area Network (\textbf{CAN}) bus chip,
for communication with 2 ELMO motion controllers (Gold Solo Whistle). 
These motor controllers receive the actual positions $(\theta^a_{pk},\, \theta^a_{pa})$ and velocities $(\dot{\theta}_{pk},\, \dot{\theta}_{pa})$ 
of the joints from an incremental encoder on the motor side of each joint and send 
these to the BBB along with the actual torque $(\tau^a_{pk}, \tau^a_{pa})$ of each joint. 
Based on these measurements, the BBB calculates the desired torques $(\tau^d_{pk}, \tau^d_{pa})$ for each joint and sends these 
commands to the ELMO motion controllers. The ELMO motion controllers send current to the motors.

\subsection{EdgeDRNN Implementation}
The EdgeDRNN runs on a \$90 entry-level MiniZed~\cite{minized}
development board. 
Fig.~\ref{fig:ampro_hardware} shows the implementation of EdgeDRNN on the MiniZed development board, which has a Zynq-7007S system-on-chip and 512\,MB off-chip DDR3L memory.
The Zynq-7007S has a single-core ARM Cortex-A9 CPU in the programming system (\textbf{PS}) and a programmable logic (\textbf{PL}) in the same package. 
EdgeDRNN is implemented on the PL with weights coming from DDR3L memory.
All modules on the PL are globally driven by a 100 MHz clock. 

\subsection{System Integration}

\begin{figure}[!t]
	\centering
	\includegraphics[width=0.47\textwidth]{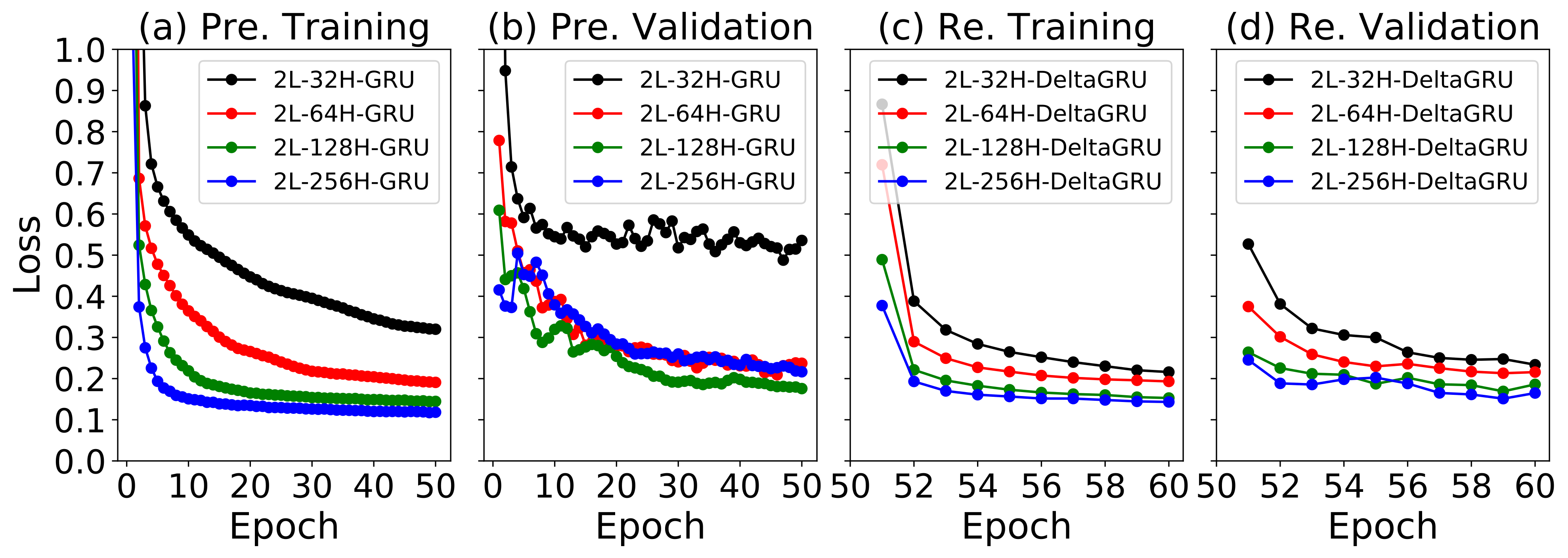}
	\caption{Training (a) and validation (b) losses of networks in RNN pretrain (epochs 1-50) and in DeltaRNN retrain (c,d) (epochs 51-60) with $M=32,64,128,256$ and $T=100$ time steps}
	\label{fig:pretrain_loss}
\end{figure}


Fig.~\ref{fig:system_photo} shows the integrated system. The MiniZed is powered by a portable USB power bank and attached to AMPRO3 opposite the BBB. To control AMPRO3, the MiniZed takes sensor data from the BBB and sends RNN controller outputs back to the BBB. An SPI bus with BBB master interfaces the two devices. 
Since the BBB samples sensor data at 200\,Hz, maintaining the latency of RNN computation under 5\,ms is crucial to ensure that the control signals are 
received by the motors on time.


\section{Results and Discussion}
\begin{table}[!t]
    \renewcommand{\arraystretch}{1.3}
    \caption{Loss of pretrained (PRE) GRU-RNN networks and retrained (RE) DeltaGRU-RNN evaluated on the validation and test sets. Number of network parameters and the epoch where the lowest validation loss is achieved are given.}
    \label{tab:pretrain_retrain}
    \centering
    \begin{tabular}{|l|c|c|c|c|}
        \hline
                  & \textbf{\#Param.} & \textbf{Epoch}  & $\mathbf{\mathcal{L}}$\textbf{-Val.} & $\mathbf{\mathcal{L}}$\textbf{-Test}\\
        \hline
        \textbf{2L-32H-GRU-PRE}      & 10 K  & 47 & 0.4878 & 0.5037  \\
        \hline
        \textbf{2L-64H-GRU-PRE}      & 38 K  & 46 & 0.2093 & 0.1900 \\
        \hline
        \textbf{2L-128H-GRU-PRE}     & 149 K & 50 & 0.1757 & 0.1677 \\
        \hline
        \textbf{2L-256H-GRU-PRE}     & 594 K & 50 & 0.2166 & 0.2120  \\
        \hline
        \textbf{2L-32H-DeltaGRU-RE}  & 10 K  & 60 & 0.2338 & 0.2523 \\
        \hline
        \textbf{2L-64H-DeltaGRU-RE}  & 38 K  & 59 & 0.2131 & 0.2207 \\
        \hline
        \textbf{2L-128H-DeltaGRU-RE} & 149 K & 59 & 0.1690 & 0.1731 \\
        \hline
        \textbf{2L-256H-DeltaGRU-RE} & 594 K & 59 & 0.1512 & 0.1689 \\
        \hline
    \end{tabular}
        \vspace{-0.7cm}
\end{table}

In the following experiments, we use the RNN hardware to perform direct control on the prosthesis in place of the PD controller. This section describes the experimental set-up for the RNN controller, the evaluation methods, and results.

\subsection{Experimental Set-up}
Four different 2-layer RNN controller networks with different layer sizes $M=32,64,128,256$ were trained with a sequence length of $T=100$ time steps, corresponding to 0.5\,s of data. 
Each network is pre-trained on cuDNN GRU layers with a learning rate of 5e-4 and batch size of 32 for 50 epochs. 
Then the cuDNN GRU parameters are loaded into DeltaGRU layers to be retrained for another 10 epochs with a learning rate of 1e-3 and batch size of 64. 
We use a pretrain-retrain approach to reduce the total training time since the cuDNN GRU is highly optimized on GPUs.
During retrain with a higher learning rate and larger batch size, GRU networks can be adapted to DeltaGRU networks with fewer epochs.
The delta threshold $\Theta$ of DeltaGRU input and hidden activations are respectively $\Theta_ x=2^{2}/2^{8}$ and $\Theta_ h=2^{7}/2^{8}$. 
We used the \texttt{Adam} optimizer and the best network
was selected according to the lowest L1 loss on the validation set. 
The network prediction performance in terms of L1 loss was evaluated on the test set. 
Training and evaluation programs are coded in PyTorch 1.2.0 with CUDA 10.1 and cuDNN 7.6.
For online evaluation, the RNN controller is used for a subject to walk on flat, uphill and downhill slopes (with $\approx 2.5^{\circ}$ slope) for 1\,min tests.
\begin{figure}[!t]
	\centering
	\includegraphics[width=0.4\textwidth]{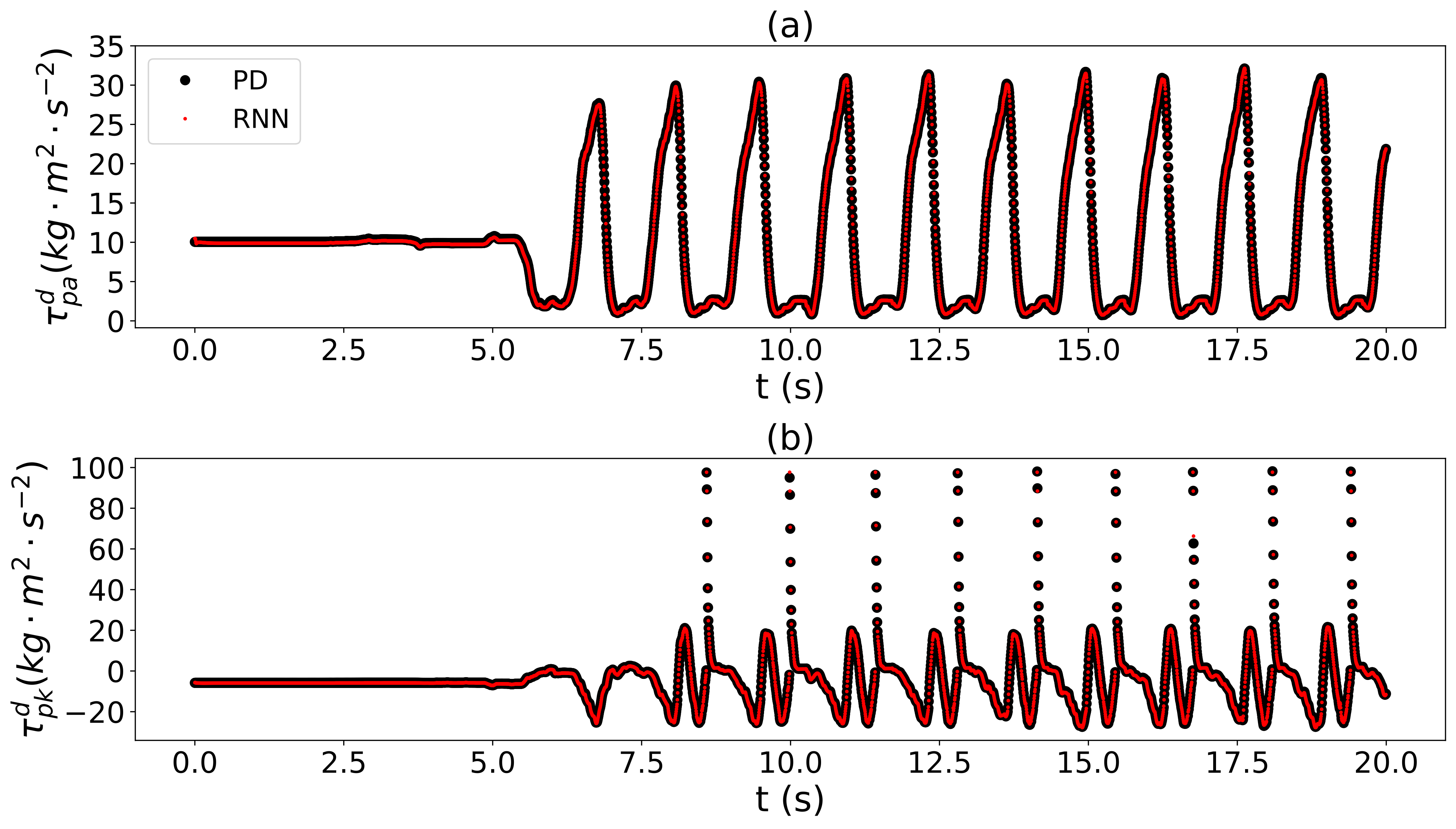}
	\caption{Comparison of $\tau^{d}_{pa}$ (a) and $\tau^{d}_{pk}$ (b) from the PD and the RNN controller on the first 20 seconds of the test set.}
	\label{fig:eval}
	    \vspace{-0.6cm}
\end{figure}

\subsection{Offline Evaluation}

\begin{figure*}[!t]
	\centering
	\includegraphics[width=0.91\textwidth]{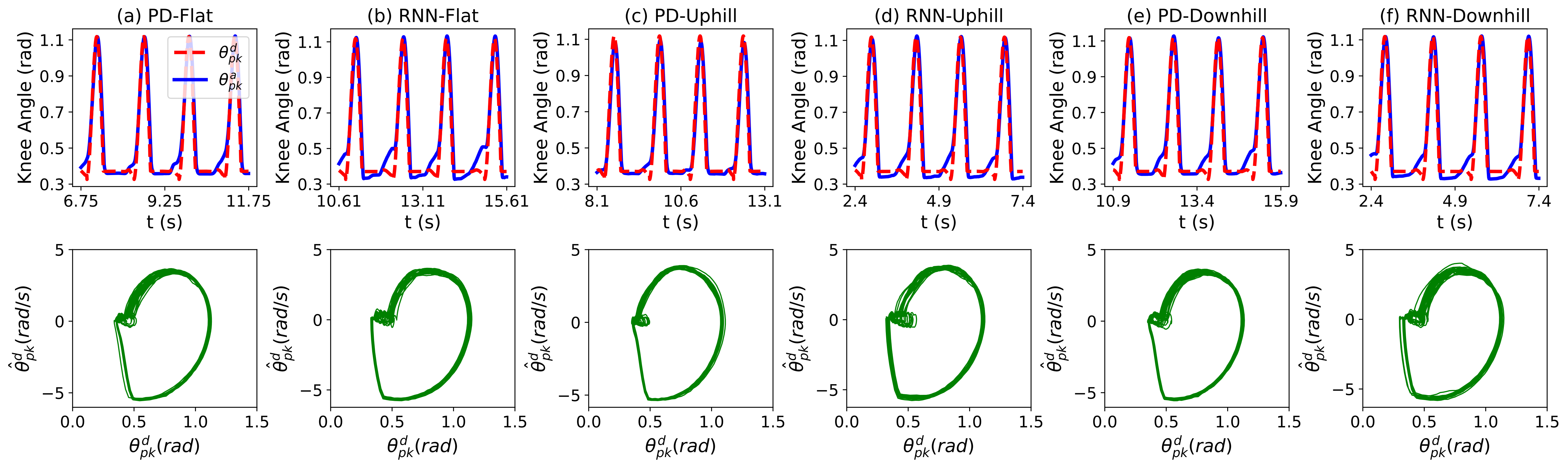}
	\caption{Experimental tracking performance (top) and phase portrait (bottom) of the prosthetic knee joint for 1\,min of walking on flat, uphill slope and downhill slope surfaces using PD and RNN controllers.}
	\label{fig:track}
    \vspace{-0.4cm}
\end{figure*}

Fig.~\ref{fig:pretrain_loss}(a) and (b) show the training and validation loss of the RNN controller network during epochs $1-50$ in the pretrain stage. 
Training losses 
decrease throughout the 50 epochs. 
Generally larger RNN layers achieve lower validation loss but the drop rate of loss decreases with increasing layer size except that we see overfitting of the 2L-256H-GRU network on the training set; it has an even higher validation loss than 2L-128H-GRU after 50 epochs.
At the beginning of the retrain stage as shown in Figs.~\ref{fig:pretrain_loss}(c) and (d), 
both training and validation losses abruptly drop due to the introduced delta threshold in DeltaGRU layers. 
After 10 epochs, the DeltaGRU-based networks reach similar validation losses as the corresponding GRU networks.
Table~\ref{tab:pretrain_retrain} shows the results of the pretrained and retrained networks.
Although the lowest test loss is achieved by the 2L-256H-DeltaGRU, we selected the 2L-128H-DeltaGRU to make the network run faster.
In rest of this paper, the RNN controller refers to this network configuration.

Fig.~\ref{fig:eval} shows the ankle and knee torque outputs of the RNN controller evaluated on the first 20\,s of the test set. 
The RNN controller outputs almost overlap with the PD controller outputs indicating good cloning of the periodic behavior.


\subsection{Online Evaluation on AMPRO3}
The RNN controller network is implemented on the prosthesis in real-time, replacing the PD controller. The RNN achieves similar tracking performance as the PD controller and generalizes well to data outside of its training set.

\newsec{Tracking Performance.}
Fig.~\ref{fig:track} shows the tracking performance (in 5\,s windows) and phase portraits of the prosthetic knee joint using both controllers for 1\,min of walking on flat, uphill and downhill slopes. Table~\ref{tab:rmse} further shows the root mean square error (RMSE) between the actual and desired trajectories of the knee and the ankle joints during the 1\,min online test on AMPRO3.
This figure shows that both controllers perform similarly regarding stability of limit cycles though the RNN controller show slightly worse tracking performance at the beginning of each knee angle spike.
The RNN controller tracking accuracy is similar to the PD controller on both knee and ankle joints for all 6 test slopes.
The results show that the RNN performed similarly to the PD controller when it encountered slopes.

\begin{table}[!t]
    \renewcommand{\arraystretch}{1.1}
    \caption{RMSE between actual and desired trajectories of PD and RNN controllers 
    during a 1\,min online test on AMPRO3.}
    \label{tab:rmse}
    \centering
    \begin{tabular}{|l|cc|cc|}
        \hline
        \textbf{Slope} & \textbf{PD-Knee} & \textbf{RNN-Knee} & \textbf{PD-Ankle} & \textbf{RNN-Ankle} \\
        \hline
        Flat (train)  & 0.0514  & 0.0592   & 0.1741   & 0.1612\\
        \hline
        Uphill        & 0.0437  & 0.0511   & 0.1563   & 0.1577\\
        \hline
        Downhill      & 0.0565  & 0.0583   & 0.1680   & 0.1708\\
        \hline
    \end{tabular}
    \vspace{-0.2cm}
\end{table}

\newsec{Generalization.}
\begin{figure}[!t]
	\centering
	\includegraphics[width=0.45\textwidth]{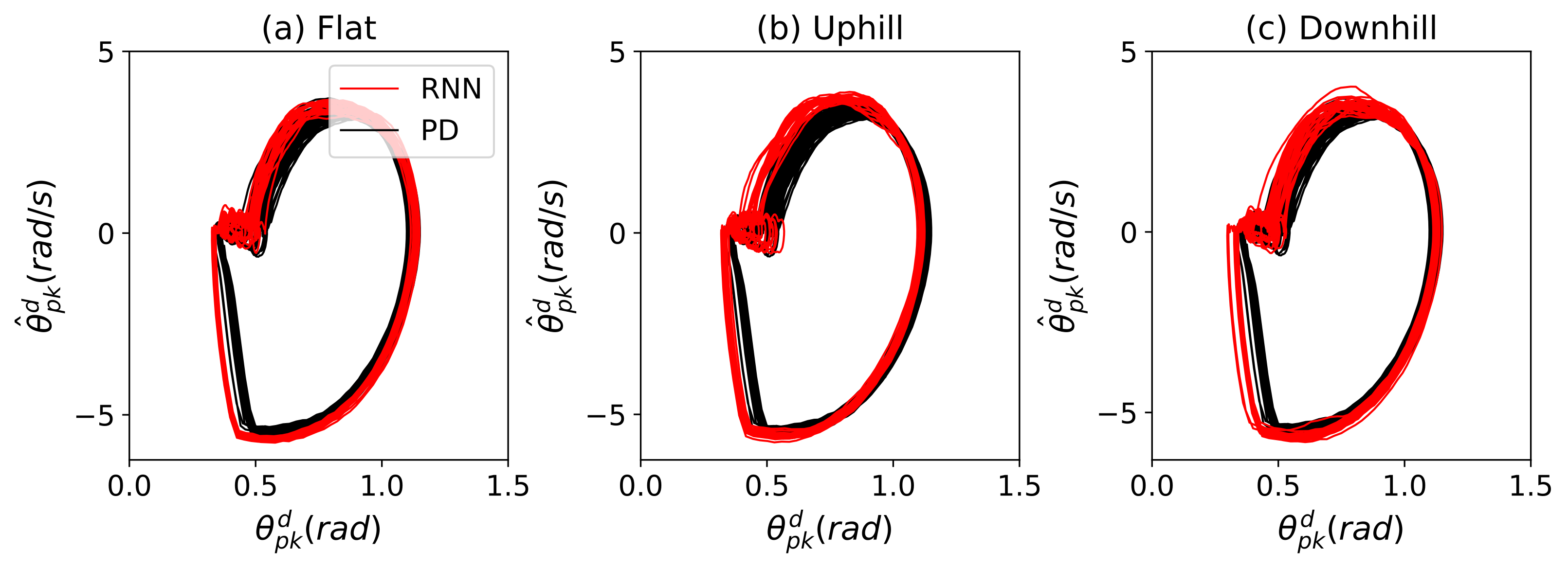}
	\caption{Comparison between the flat-ground training set trajectories from the PD controller (identical in each plot)
	and trajectories from the RNN controller on different slopes during a 1\,min online test on AMPRO3.}
	\label{fig:comp}
    \vspace{-0.9cm}
\end{figure}
To visualize the generalization ability of the RNN trained only on flat walking, Fig.~\ref{fig:comp} compares 
trajectories recorded during online tests of the RNN controller on flat and sloped walking with training set trajectories on flat ground.
Fig.~\ref{fig:comp}(a) shows that RNN and the PD trajectories 
on flat ground are nearly indistinguishable, which makes
sense because the RNN is trained on this data.
Figs.~\ref{fig:comp}(b) and (c) 
show that the RNN maintains stable trajectories in sloped walking, which is outside the training data. This along with the results of Table~\ref{tab:rmse} suggest the RNN generalized well.


\subsection{Hardware Latency \& Power Consumption}
\begin{table}[!t]
    \renewcommand{\arraystretch}{1.1}
    \caption{System power breakdown (*measured by a USB power meter)}
    \label{tab:power}
    \centering
    \begin{tabular}{|l|c|}
        \hline
                                    & Power (W)  \\
        \hline
        \textbf{PS (ARM) Dynamic}        & 1.312 W   \\
        \hline
        \textbf{PL (EdgeDRNN) Dynamic}       & 0.123 W   \\
        \hline
        \textbf{On-chip Static}     & 0.124 W   \\
        \hline
        \textbf{Off-chip (DDR3, storage, I/O, etc.)}           & 0.453 W   \\
        \hline
        \textbf{MiniZed (Total)*}           & 2.012 W    \\
        \hline
    \end{tabular}
    \vspace{-0.5cm}
\end{table}

We measured the latency of computing the 2L-128H-DeltaGRU on EdgeDRNN and BBB using a 1\,min segment of the test set. 
Each sample requires about 300k Op (two times the number of RNN weights).
The average latency of EdgeDRNN is 20.9\,$\mu$s 
(ranging from 9 to 140 $\mu$s) 
while the latency of the BBB is 1062.7 $\mu$s 
(ranging from 288 to 3685 $\mu$s).
Thus EdgeDRNN achieves around 51x lower latency than the BBB.
Table~\ref{tab:power} show the power breakdown of the MiniZed system. Whole system power is 2.012\,W while the EdgeDRNN consumes only 0.123\,W.
Thus the EdgeDRNN system achieves 7\,GOp/s/W total and 116\,GOp/s/W incremental power efficiency.

\section{Conclusion}
This paper presents the first end-to-end RNN solution for controlling a transfemoral prostheses in a continuous action space. Experimental results show that the DeltaGRU network can be properly trained to replace the PD controller with similar tracking performance. It also maintains stable trajectories on slopes, indicating it generalized to data that is not part of its training set. With only 2\,W power consumption, the EdgeDRNN accelerator on MiniZed is capable of sub-millisecond latency inference of large GRU-RNNs and realizes the prosthesis control in real-time. 
Future work is needed to study the stability of the RNN controller on extended test conditions (e.g. different terrains) and on different subjects. 
Although this RNN design does not improve upon the PD controller, its success in emulating the PD controller in real-time opens the possibility of training an RNN to learn more complex controllers that are computationally expensive and implement the networks on hardware accelerators 
for lower latency and energy consumption of the system. This hardware demonstration is a first step towards using RNNs to synergistically incorporate complex individual human gait behavior into prosthesis control, thereby closing the loop between human intent and human-prosthesis walking.






\bibliographystyle{IEEEtran}
\balance
\bibliography{IEEEabrv}


\end{document}